\documentclass[aps,pra,twocolumn,superscriptaddress,
floatfix]{revtex4-1}
\usepackage{amsmath}
\usepackage[next]{inputenc}
\usepackage[dvips]{epsfig}
\usepackage{bbm,bm,bbold}
\usepackage{booktabs}
\usepackage{svg} 
\usepackage{multirow}
\usepackage{hhline}
\usepackage{comment}
\usepackage{textcomp}
\usepackage{float}
\usepackage{enumitem} 

\usepackage{amsmath,amsfonts,amssymb,amsthm}
\usepackage{color}
\usepackage{graphicx,latexsym}
\usepackage{svg} 

\usepackage[colorlinks=true,urlcolor=blue,citecolor=blue,linkcolor=blue]{hyperref}

\usepackage{lipsum}

\usepackage{nameref}
\usepackage{varioref}
\usepackage{cleveref}

\usepackage{natbib}

\begin{document}
\renewcommand{\vec}{\mathbf}
\renewcommand{\Re}{\mathop{\mathrm{Re}}\nolimits}
\renewcommand{\Im}{\mathop{\mathrm{Im}}\nolimits}
\newcommand\scalemath[2]{\scalebox{#1}
{\mbox{\ensuremath{\displaystyle #2}}}}

\title{Weyl magnetoplasma waves in magnetic Weyl semimetals}

\author{Yuanzhao Wang}
\affiliation{School of Physics and Astronomy, Monash University, Victoria 3800, Australia}

\author{Oleg V. Kotov}
\affiliation{Departamento de Fisica Teorica de la Materia Condensada, Universidad Autonoma de Madrid, E-28049 Madrid, Spain
}

\affiliation{
Condensed Matter Physics Center (IFIMAC), Universidad Aut\'onoma de Madrid, E-28049 Madrid, Spain}

\author{Dmitry K. Efimkin}
\email{dmitry.efimkin@monash.edu}
\affiliation{School of Physics and Astronomy, Monash University, Victoria 3800, Australia}

\begin{abstract}
Weyl degeneracies in spectra of magnetoplasma waves enable nonreciprocal energy flow and topologically protected modes, yet conventional materials require impractical magnetic fields to operate. Developing an effective Hamiltonian framework for magnetic Weyl semimetals, we show that these systems overcome the limit, hosting Weyl magnetoplasma physics at zero field due to their giant intrinsic anomalous Hall response. The resulting topology supports nonreciprocal modes localized at magnetic domain walls, including a pair of topological "Fermi-arc-like" modes and additional bound states. These effects are fully developed across a broad THz window, and we propose feasible experimental routes for their detection.
\end{abstract}

\date{\today}
\maketitle

\noindent\textcolor{blue}{\emph{Introduction}}.
Magnetic Weyl semimetals have emerged as a fertile platform for topological phenomena, including the chiral anomaly~\cite{BurkovIOPcmeReview}, large anomalous Hall responses~\cite{BurkovAHE}, surface Fermi-arc transport~\cite{FermiArcTrasnport1,FermiArcTrasnport2,FermiArcTrasnport3} (See Refs.~\cite{Armitage:RMP2018, ZHasan:reviewDiscovery,burkov2018weyl} for a review). Their magnetic order further gives rise to distinctive electrodynamic effects, such as natural optical activity~\cite{FaradeyKerr1,FaradeyKerr2} and nonreciprocal surface plasmon-polariton modes~\cite{SPPWeyl1,SPPWeyl2,SPPWeyl4,SPPWeyl5}. Closely related Weyl-type excitations have recently been realised in a variety of non-electronic settings, including photonic and phononic metamaterials~\cite{WeylPhotonic,WeylPhotonicAcoustic}, and magnetised plasmas~\cite{PlasmaTop1}.

The spectrum of magnetoplasma (MP) waves --- coupled matter and electromagnetic waves supported by an electron gas in a magnetic field --- is split, yet retains degeneracies between different branches that have only recently been identified as topological Weyl nodes~\cite{PlasmaTop1,PlasmaTop2,PlasmaTop3}. These topological degeneracies enable protected trapped modes and nonreciprocal energy transport, placing MP waves among the earliest continuum platforms for topological photonics~\cite{TopPhotonicsReview,GyrotropicInterfacesReview}. In practice, however, the Hall-induced interbranch splitting achievable with laboratory magnetic fields remains small, even in semiconductors with ultralight carriers such as InSb, severely limiting the accessible frequency window and hindering experimental observation~\cite{PlasmaTopExp}. By contrast, magnetic Weyl semimetals, including Co$_3$Sn$_2$S$_2$, Co$_2$MnGa, and MnBi$_2$Te$_4$, exhibit a giant anomalous Hall response in zero magnetic field~\cite{AHEMain1,AHEMain2,AHESecondary1,AHESecondary2,AHESecondary3}, raising the question of whether this intrinsic Hall effect alone can generate Weyl MP physics.

\begin{figure}[t]
\vspace{-0.05in}
\centering
\includegraphics[scale=0.5]{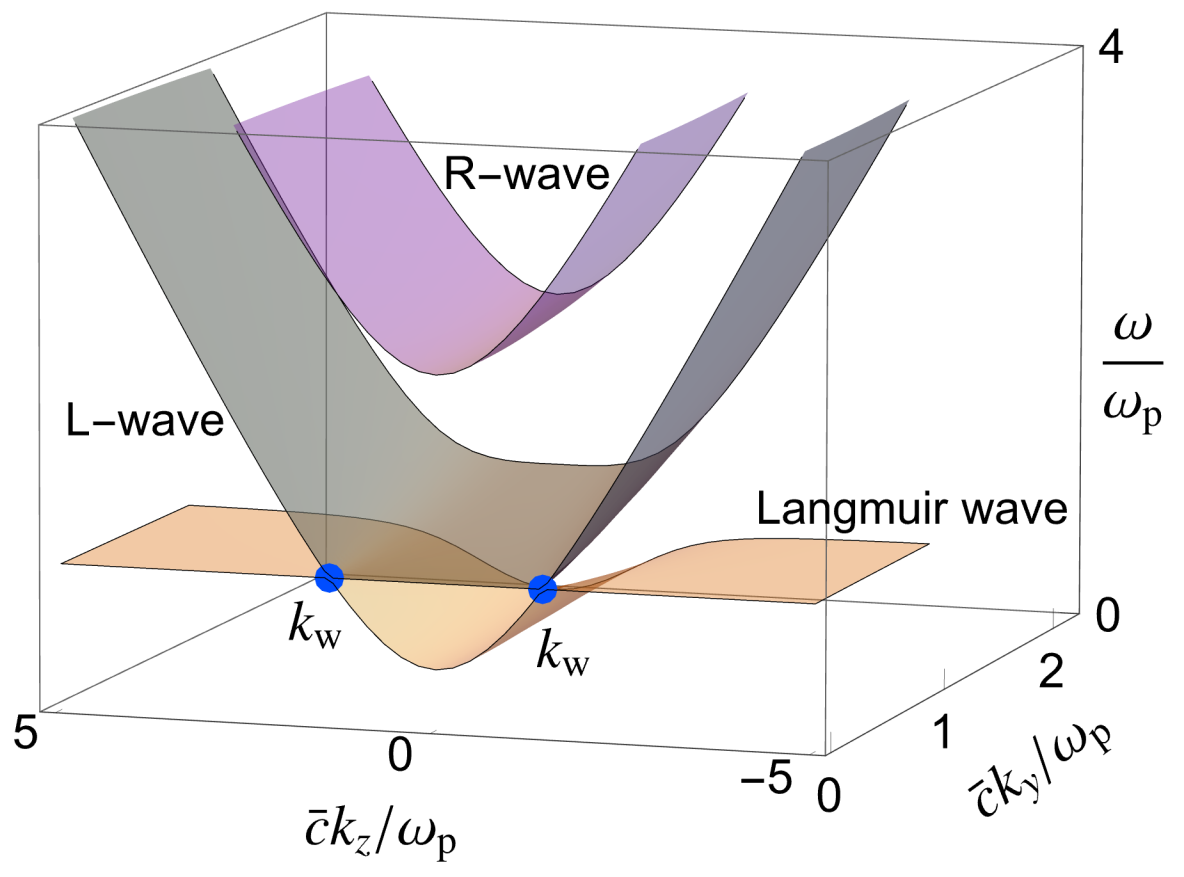}
\vspace{-0.06in}
\caption{Dispersion of magnetoplasma waves in a magnetic Weyl semimetal, comprising longitudinal Langmuir (plasma) oscillations and two transverse circularly polarized modes of left- (L) and right-handed (R) helicity. These polarizations are well defined only for propagation along the gyrotropic axis ($z$), where branch decoupling occurs and a pair of Weyl nodes (blue dots) is formed.} \vspace{-0.20in}\label{BulkDispersion}
\end{figure}

In this Letter, we re-examine the dynamics of magnetoplasma (MP) waves in magnetic Weyl semimetals within the effective Hamiltonian framework~\cite{PlasmaTop1,PlasmaTop2,PlasmaTop3}. This approach enables a topological classification of the wave spectrum and reveals that the intrinsic anomalous Hall response of the medium generates a pair of Weyl nodes, as shown in Fig.~\ref{BulkDispersion}. Moreover, the framework naturally extends to spatially nonuniform systems and reveals that, despite the absence of a full bulk gap --- restricting standard bulk-edge correspondence arguments --- the underlying Weyl character of the spectrum still gives rise to unconventional domain-wall-localized modes, both topological and nontopological. For Bloch-type DWs, these modes closely emulate Fermi-arc states. By contrast, modes trapped at N\'eel-type DWs do not exhibit Fermi-arc-like characteristics, yet remain strongly nonreciprocal. Finally, we discuss experimentally feasible routes for detecting the predicted modes.

\noindent\textcolor{blue}{\emph{Bulk magnetoplasma waves}}. For concreteness, we consider a minimal model of a magnetic Weyl semimetal that preserves inversion symmetry and hosts 
$N$ degenerate pairs of Weyl nodes separated in momentum space, but without an energy splitting. The coupled dynamics of fields and matter are then governed by Maxwell's equations 
\begin{subequations}
\begin{align}
\label{Maxwell1}
\mathrm{curl}\, \vec{H}(\vec{r},t) & = \frac{4\pi}{c} \left[\vec{j}_\mathrm{D}(\vec{r},t)+\vec{j}_\mathrm{A}(\vec{r},t) \right]+\frac{1}{c} \partial_t \vec{D} (\vec{r},t), \\
\mathrm{curl}\, \vec{E}(\vec{r},t) & = -\frac{1}{c} \frac{\partial \vec{H} (\vec{r},t)}{\partial t},
\label{Maxwell2}
\end{align}
complemented by the matter response
\begin{equation}
\frac{\partial\vec{j}_\mathrm{D}(\vec{r},t)}{\partial t}=2N\mathcal{D} \vec{E}(\vec{r},t), \quad \;  \vec{j}_\mathrm{A}(\vec{r},t) = \frac{N e^2}{2\pi^2 \hbar}\,[\mathbf{b}\times\mathbf{E}].
\label{MatterResponse}
\end{equation}
\end{subequations} A distinctive feature of Weyl semimetals is that the electric current comprises two physically distinct contributions. The first term, $\vec{j}_{\mathrm D}(\vec{r},t)$, originates from the electric-field-induced shift of the electron distribution within the Fermi seas surrounding the Weyl nodes. This response is captured by the Drude model with weight $\mathcal{D}=\nu_{\mathrm F} v^{2} e^{2}/3$, where $\nu_{\mathrm F}=\epsilon_{\mathrm F}^{2}/2\pi^{2}(\hbar v)^{2}$ is the density of electronic states at the Fermi energy $\epsilon_{\mathrm F}$ and $v$ is the electron velocity. The second contribution, $\vec{j}_\mathrm{A}(\vec{r},t)$, represents the topological 
Anomalous Hall response~\cite{burkov2018weyl}, which is determined solely by the separation between two
Weyl nodes in momentum space $2\vec{b}$~\footnote{It should be noted that the anomalous Hall response can also be described by axionic electrodynamics with the gradient of the axion field $\nabla \theta(\vec{e})=2\vec{b}$; see the Supplemental
Material for an extended discussion~\cite{SM}}.

The conventional approach to dealing with the system of Eqs.~(\ref{Maxwell1}-\ref{MatterResponse}) involves the exclusion of the oscillating magnetic field and the electric current, 
which results in the Helmholtz equation:
\begin{equation}
\label{WaveEquation}
\nabla\times(\nabla \times \vec{E})=\frac{\omega^2}{\bar{c}^2}\frac{\hat{\varepsilon}(\omega)}{\varepsilon_\infty} \vec{E}.
\end{equation}
Here, we have incorporated the high-frequency response of atomic electronic shells via the optical dielectric constant $\epsilon_\infty$ as $\vec{D}(\vec{r},t)=\epsilon_\infty \vec{E}(\vec{r},t)$. It defines the
 speed of electromgnatic waves in the media $\bar{c}=c/\sqrt{\varepsilon_\infty}$ and its dielectric tensor $\hat{\varepsilon}(\omega)$ is given by
\begin{equation}\label{dielectric tensor}
\frac{\hat{\varepsilon}(\omega)}{\varepsilon_\infty} =
\begin{pmatrix}
\varepsilon_\mathrm{L} & -i g_z  & i g_y \\
i g_z & \varepsilon_\mathrm{L} & -i g_x \\
-i g_y & ig_x & \varepsilon_\mathrm{L}
\end{pmatrix}
\end{equation}
Here, the longitudinal component $\varepsilon_\mathrm{L}(\omega)=1-\omega_\mathrm{p}^{2}/\omega^{2}$ follows the lossless Drude model with plasma frequency $\omega_{\mathrm{p}}=\sqrt{8\pi N D/\epsilon_\infty} $. In contrast, the off-diagonal elements are determined solely by the anomalous Hall response. The corresponding gyration vector 
\begin{equation}
\vec{g}= \frac{c_\mathrm{b} \vec{b}}{\omega} \quad \quad c_\mathrm{b}=\frac{2 e^2}{\pi\hbar\varepsilon_\infty}
\end{equation}
is directed along the momentum separating the Weyl nodes, and its magnitude is characterized by the frequency $\omega_\mathrm{b}=c_\mathrm{b} b$. 

The conventional approach is well-suited for analyzing a broad class of electrodynamic phenomena, including natural optical activity resulting in Faraday and Kerr effects~\cite{FaradeyKerr1,FaradeyKerr2}, and nonreciprocal plasmon-polaritons localized Weyl semimetals surfaces~\cite{SPPWeyl1,SPPWeyl2,SPPWeyl4,SPPWeyl5}, waveguides and closed geometries~\cite{WeylSemimetalWaveGuide,WeylSemimetalSphere}, and Weyl semimetal-based gyrotropic interfaces~\cite{WeylSemimetalGyrotropic} (See Ref.~\cite{WeylSemimetalLightControlReview} for a Review). However, it overlooks the nontrivial topological properties of the bulk modes. These features become apparent only when we reformulate the classical wave equations, Eqs.~(\ref{Maxwell1}-\ref{MatterResponse}),  in terms of an effective Hamiltonian, following the framework developed for magnetised plasma~\cite{PlasmaTop1,PlasmaTop2,PlasmaTop3,EfimkinHelicon} and other systems~\cite{MagnetoPlasmons2D,EfimkinMP,MagnetoplasmonZubin, TopologyEquatorialWaves,EfimkinEffHSF, EfimkinPlasmonScattering,EfimkinPlasmonScatteringAnomalous}. If we introduce the frequency-wave vector domain, $\frac{\partial}{\partial t} \rightarrow -i\omega$ and $\nabla \rightarrow i\mathbf{k}$, the system of equations, Eqs.~(\ref{Maxwell1}-\ref{MatterResponse}), can be presented as eigenvalue problem $\omega \psi_\mathrm{pl} = \hat{\mathcal{H}}_\mathrm{pl}(\vec{k}) \psi_\mathrm{pl}$. Here, 9-component   vector  $\psi_\mathrm{pl} = \{\sqrt{\varepsilon_{\infty}}\vec{E}(\vec{k}),\, \vec{H}(\vec{k}),\, 4\pi\vec{j}_\mathrm{D}(\vec{k})/(\omega_{\mathrm{p}}\sqrt{\varepsilon_{\infty}})\}$ serves as the state vector, and the matrix $\hat{\mathcal{H}}_\mathrm{pl}$ --- interpreted as the effective Hamiltonian of the system --- is given by 
\begin{equation}
\label{HamiltonianPlasma}
\hat{\mathcal{H}}_\mathrm{pl}(\vec{k})=\begin{pmatrix}
-i c_\mathrm{b} \hat{K}_\vec{b} & - \bar{c} \hat{K}_{\vec{k}} & - i \omega_\mathrm{p} \hat{1}\\
\bar{c} \hat{K}_{\vec{k}} & 0 & 0 \\
i \omega_\mathrm{p} \hat{1} & 0 &  0   \\
\end{pmatrix}.
\end{equation}
Here, the matrix $\hat{K}_{\vec{k}}$ represents the action of
$-i\,\mathrm{curl}$,
and its antisymmetric structure~\footnote{The explicit form of the antisymmetric operator  $\hat{K}_{\vec{k}}$ is $$\hat{K}_{\vec{k}}= \begin{pmatrix} 0 & - k_z & k_y \\ k_z & 0 &-k_x \\ -k_y & k_x & 0\\ \end{pmatrix}$$.} ensures the Hermiticity of the effective Hamiltonian
$\hat{\mathcal H}_{\mathrm{pl}}(\mathbf k)$. The resulting eigenvalues and eigenfunctions fully characterize the dispersion relations and topological properties of bulk modes. 

\noindent\textcolor{blue}{\emph{Bulk modes and their nontrivial topologies}:} 
The dispersion relations for bulk MP modes are determined by the positive eigenvalues of $\hat{\mathcal{H}}_\mathrm{pl}(\vec{k})$. Without loss of generality, we set $\vec{b}=b\,\vec{e}_z$, in which case the bulk spectrum shown in Fig.~\ref{BulkDispersion} consists of three branches --- longitudinal Langmuir waves (plasma) oscillations) and the two transverse circularly polarised left- (L-) and right-handed (R-) modes --- each with a gap on the order of the plasma frequency~$\omega_\mathrm{p}$. In contrast to conventional magnetised plasmas, the spectrum conspicuously lacks the low-frequency helicon branch (also known as whistlers~\cite{PlasmaBook}).

As evident in Fig.~\ref{BulkDispersion}, the dispersion curves of the L and Langmuir modes intersect at two singular points in reciprocal space. These crossings occur at the plasma frequency $\omega_{\mathrm p}$ and at wave vectors $\pm k_{\mathrm W}\,\vec{e}_{z}$ of magnitude $k_{\mathrm W}=\sqrt{\omega_{\mathrm b}\,\omega_{\mathrm p}}/\bar{c}$, aligned with the direction of~$\vec{b}$. Only for propagation along these special directions do the L and Langmuir modes reduce to purely transverse circularly polarised and longitudinal waves, ensuring that the associated current and electromagnetic field oscillations remain independent and fully decoupled. The resulting linear band crossings possess the characteristic structure of Weyl nodes, as confirmed by the topological analysis.

The topology of the MP  modes is characterized by the distribution of the Berry curvature vector 
\begin{equation}
\vec{F}^\alpha_n(\vec{k}) =
2 \varepsilon_{\alpha\beta \gamma} \,\mathrm{Im} \sum_{m \neq n} \frac{
\left\langle n \left| \frac{\partial H}{\partial k_\beta} \right| m \right\rangle
\left\langle m \left| \frac{\partial H}{\partial k_\gamma} \right| n \right\rangle
}{
(\omega_n - \omega_m)^2
}
\end{equation}
across the reciprocal space. First, the Berry flux across any closed surface surrounding any of two Weyl nodes is quantized as $C_n^\pm=\frac{1}{2\pi}\int_S \vec{F}_n d\vec{S}=\mp 1$ for L- and Langmuir waves, accordingly, as anticipated for Weyl nodes. Second, the Weyl topology can be linked with the Chern numbers characterizing Berry phase flux across the reciprocal space sections perpendicular to the vector $\vec{b}$ as
\begin{equation}
\label{Chern number}
C_n(k_z)= \frac{1}{2\pi} 
\int\!\!\int dk_x\, dk_y\,
F^z_n(k_x,k_y).
\end{equation}
An explicit evaluation (see the Supplemental Material (SM) for details~\cite{SM}) shows that the set of Chern numbers for reciprocal-space sections between the Weyl nodes is $(1,\,0,\,-1)$, whereas for sections outside the nodes it becomes $(1,\,-1,\,0)$. As anticipated, topological transitions occur when the section crosses a Weyl node.

A prominent consequence of Weyl-type topology is the possible emergence of protected surface arc states. Their robustness, however, requires the presence of a global band gap in the plane of reciprocal space normal to the surface. In the MP spectrum considered here, such a global gap is not fully developed-much like in the overdoped regime of conventional magnetised plasmas ~\cite{PlasmaTop1}. These observations naturally raise the question: to what extent can the uncovered Weyl nature of MP waves support any unconventional trapped modes?

\noindent\textcolor{blue}{\emph{The domain wall problem}}. 
In magnetic Weyl semimetals, the separation between electronic Weyl nodes of opposite chirality is locked to the underlying magnetisation. When the magnetisation reverses across a domain wall (DW), this separation vector flips, causing the chirality of Weyl nodes to interchange and favoring the emergence of domain-wall-confined electronic Fermi-arc states~\cite{FermiArc1,FermiArc2}. Because MP Weyl nodes are aligned with their electronic counterparts, they likewise swap across the wall. Topology therefore suggests the existence of arc-like trapped MP modes, although the lack of a full global gap implies that their detailed structure and even presence may depend on the details of the DW profile.

To assess whether such MP arc states are indeed present, we analyse the DW problem for both Bloch and N\'{e}el configurations. Since the separation of the Weyl nodes follows the local magnetisation, the spatial variation of the associated $\vec{b}$-vector can be parametrised as: \begin{align*}
& \text{Bloch:}  \quad  b_x=0,\quad b_y = \frac{b}{\cosh\left(\frac{x}{d}\right)}, \quad b_z=\,b \tanh\!\left(\frac{x}{d}\right) \\[6pt]
& \text{Neel: } \quad   b_x=\frac{b}{\cosh\left(\frac{x}{d}\right)},\quad b_y = 0, \quad b_z=\,b \tanh\!\left(\frac{x}{d}\right)
\end{align*}
Here, $d$ is the domain-wall width. Reformulating the problem in terms of an frequncy-independent effective Hamiltonian provides a natural and efficient framework for calculating the magnetoplasmon spectrum in spatially nonuniform problems, by computing the eigenvalues of its real-space-discretized version of $\hat{\mathcal{H}}_\mathrm{pl}(\hat{k}_x,k_y,k_z)$. These calculations, however, rely on periodic boundary conditions, which require the magnetization to return to its original state. This can be accomplished by introducing a pair of domain walls,
$\vec{b}'(x) = \vec{b}(x - L/2) - \vec{b}(x + L/2)$, 
where $L$ denotes the separation between the walls. The parameter $L$ is chosen sufficiently large to prevent overlap of the interface modes and thus avoid any distortion of the resulting spectrum. Before presenting numerical results, we first discuss the analytical solution for a simplified DW profile, which offers clear insight into the formation of the interface states and their topological origin.

 \noindent\textcolor{blue}{\emph{Analytical solution for a simplified DW profile}}. For realistic DW profiles, the gyration vector magnitude remains approximately uniform, while its direction rotates smoothly between $\pm \vec{e}_{z}$. For simplicity, we retain only its $\vec{e}_{z}$ component and approximate the DW profile as
\begin{equation}
b_{x}=0, \qquad 
b_{y}=0, \qquad
b_{z}=b\tanh\!\left(\frac{x}{d}\right).
\label{simplifiedDW}
\end{equation}
If we further restrict ourselves to the Voigt configuration, where the wave vector is perpendicular to the gyration vector, $\vec{k}=k_{y}\,\vec{e}_{y}$, the problem becomes analytically tractable~\cite{Raikh}. Substituting Eq.~(\ref{simplifiedDW}) into the Helmholtz equation, Eq.~(\ref{WaveEquation}), results in the following system of equations for the electric field 
components:
\begin{subequations}
\begin{align}
\Big[\omega^2 \varepsilon_L - \bar{c}^2 k_y^2\Big] E_x 
+ \Big[\bar{c}^2 k_y \hat{k}_x + i \omega^2 g_z(x)\Big] E_y &= 0 \\[6pt]
\Big[\bar{c}^2 k_y \hat{k}_x - i \omega^2 g_z(x)\Big] E_x 
+ \Big[\omega^2 \varepsilon_L - \bar{c}^2 \hat{k}_x^{\,2}\Big] E_y &= 0 \\[6pt]
\Big[\bar{c}^2(\hat{k}_x^{\,2} + k_y^2) -\omega^2\varepsilon_L\Big] E_z &= 0
\end{align}
\label{Helmeholtz2}
\end{subequations}
\noindent These equations have multiple solutions, and it is instructive to describe them separately.

\begin{figure}[t]
  \centering
  \includegraphics[scale=0.45]{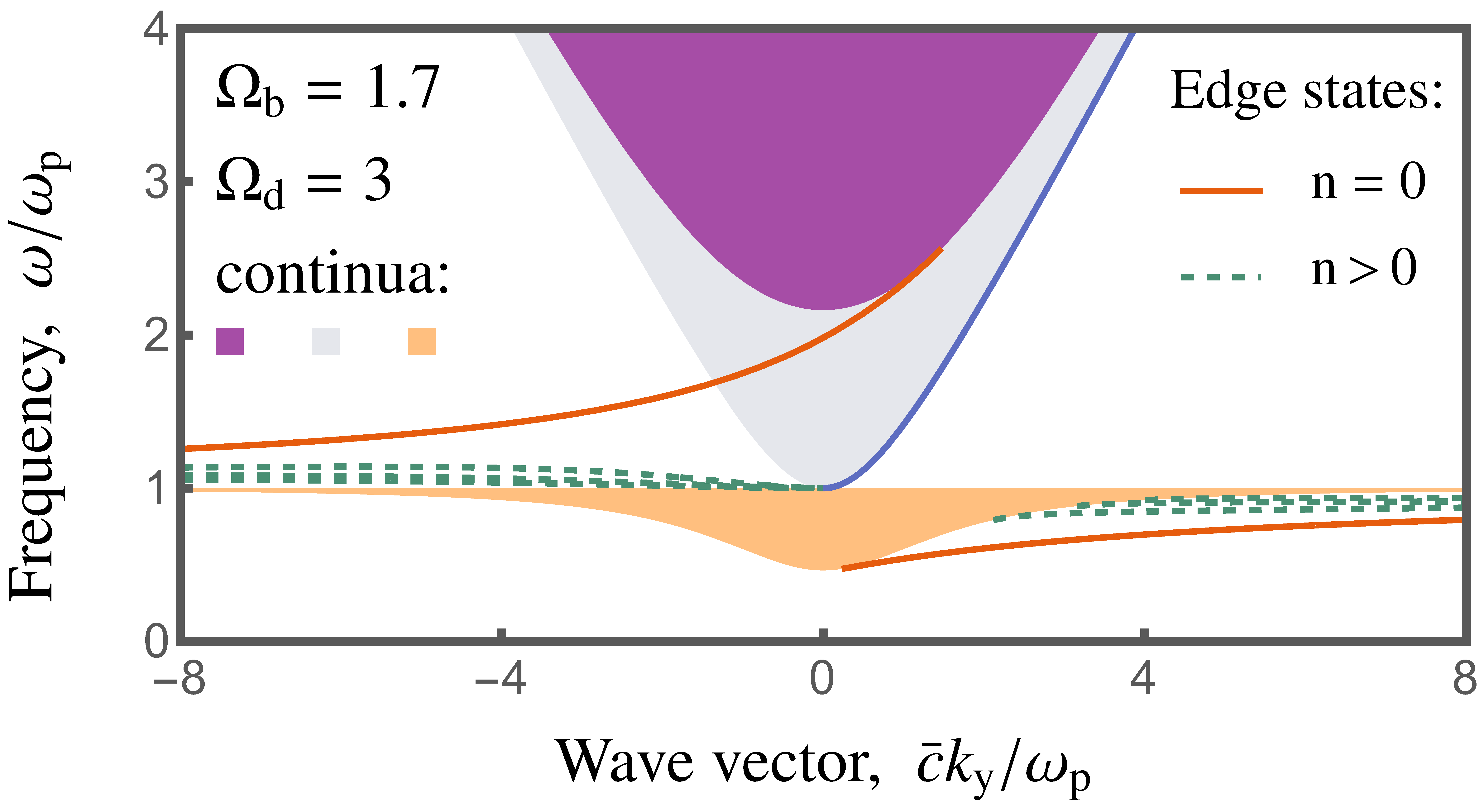}
  \vspace{-0.1in}
   \caption{Analytical spectrum of bulk (shaded regions) and DW-confined (lines) MP waves for the simplified profile in Eq.~(\ref{simplifiedDW}) at $k_z = 0$, exhibiting two topological DW modes(blue and red) consistent with bulk--edge correspondence.}
  \vspace{-0.15in}
  \label{fig:comparision}
\end{figure}

First, we notice that the oscillations of the electric field along the gyrotropy axis, $E_z$ are decoupled from the other two components and their dynamics is not affected by the anomalous Hall response. The solutions of Eq.~(\ref{Helmeholtz2})-c are, therefore, propagating plane waves and determine the bulk continuum $\omega=\sqrt{\omega_\mathrm{p}^2+ \bar{c}^2(k_x^2 + k_y^2)}$ where $k_x$ varies. It is presented in Fig.~\ref{fig:comparision}-a in light gray. 

Second, Eqs.~(\ref{Helmeholtz2})-a and b admit a special solution 
\begin{equation}\label{Jackiw-Rebbi state}
E_x(x) \sim
\exp\!\left[
-\frac{\omega^2}{\bar{c}^2 k_y}
\int^{x}_0 b_z(x')\,dx'
\right],
\qquad 
E_y = 0
\end{equation}
with the dispersion relation
$\omega = \sqrt{\omega_{\mathrm p}^{\,2} + \bar{c}^{2}k_y^{\,2}},$ which is independent of the specific domain-wall profile and is presented in Fig.~\ref{fig:comparision} as solid blue curve. The mode is normalisable and represents a genuinely trapped interface excitation only for $k_y>0$. Its existence relies on the kink-like spatial dependence of $b_z(x)$, in close analogy with the electronic Jackiw--Rebbi bound state~\cite{JackiwRebbi} or the Kelvin plasmon mode in a two-dimensional electron gas trapped at the kink of the magnetic field~\cite{MagnetoPlasmons2D,EfimkinMP,MagnetoplasmonZubin}.

Third, if $\omega\neq\sqrt{\omega_\mathrm{p}^2+ \bar{c}^2 k_y^2}$, the pair of Eqs.~(\ref{Helmeholtz2})-a and b can be combined in a single equation for $E_y$. If we use the explicit form of the domain-wall profile \( b_z(x) = b \tanh(x/d) \) and rescale the coordinate as \( x \rightarrow x/d \), the resulting eigenvalue equation is recast into a one-dimensional quantum-mechanical problem governed by the P\"oschl--Teller (PT) confining potential:
\begin{equation}\label{PT potential}
-\frac{\partial^2E_y}{\partial x^2}
-\frac{\lambda(\lambda+1)}{\cosh^2(x)}E_y=\epsilon E_y
\end{equation}
The parameter $\lambda$ and energy $\epsilon$ depend on both the wave vector $k_y$ and frequency $\omega$  and are given by
\begin{subequations}\label{parameter equations}
\begin{align}
&\lambda(\lambda + 1)
= \frac{\omega \omega_\mathrm{b}\, (\omega \omega_\mathrm{b} - \omega_\mathrm{d} \bar{c} k_y)}
{\omega_\mathrm{d}^{2}(\omega^{2} - \omega_\mathrm{p}^{2})},
 \\[8pt]
&\epsilon
= \frac{\big(\omega^{2} - \omega_\mathrm{p}^{2} - \bar{c}^{2} k_y^{2}\big)(\omega^{2} - \omega_\mathrm{p}^{2})
- \omega^{2} \omega_\mathrm{b}^{2}}
{\omega_\mathrm{d}^{2}(\omega^{2} - \omega_\mathrm{p}^{2})}
\end{align}
\end{subequations}
Here, we have introduced the frequency scale $\omega_\mathrm{d}=\bar{c}/d$, which characterizes the smoothness of the DW profile. The spectrum of the PT problem~\cite{Susy1} includes a continuum of extended states and a set of discrete states as
\begin{subequations}\label{continuous and discrete equation}
\begin{align}
\text{continuous:} \qquad  \epsilon > 0, \\[6pt]
\text{discrete:}   \qquad  \epsilon_n = -(\lambda - n)^2,\quad n \le \lambda .
\end{align}
\end{subequations}
Here, $n = 0, 1, \ldots$ is integer. The delocalized states $\epsilon > 0$ define the other two continua for bulk magnetoplasma waves, presented in Fig.~\ref{fig:comparision}-a in the purple and orange areas. In contrast, the discrete bound states of the PT problem are intricately related to the trapped MP modes. The closed equation for their dispersion can be obtained if we combine Eq.~(\ref{continuous and discrete equation}) with Eq.~(\ref{parameter equations}), and it is given by
\begin{equation}\label{smart equation}
\begin{split}
n^2+n+(2n+1)\sqrt{-\epsilon}+ &\frac{\omega \omega_\mathrm{b} \bar{c} k_y}{\omega_\mathrm{d}(\omega^2-\omega_\mathrm{p}^2)}+\\&\frac{\omega_p^2+\bar{c}^2 k_y^2-\omega^2}{\omega_\mathrm{d}^2}=0
\end{split}
\end{equation}
Its nonlinear character allows multiple solutions to exist for the same $n = 0,1,\ldots$ and $k_y$. In the dimensionless variables $\omega/\omega_{\mathrm p}$ and $\bar{c}\,k_y/\omega_{\mathrm p}$, the MP dispersion depends on two dimensionless parameters, $\Omega_{\mathrm b} = \omega_{\mathrm b}/\omega_{\mathrm p}$ and $\Omega_{\mathrm d} = \omega_{\mathrm d}/\omega_{\mathrm p}$ characterizing the magnitude of anomalous Hall response and the domain wall width. Guided by the estimates discussed below, we adopt the representative choice $\Omega_{\mathrm b}=1.7$ and $\Omega_{\mathrm d}=3$ for the figures presented here, and a more extensive analysis is presented in the SM~\cite{SM}.   The corresponding solutions of Eq.~(\ref{smart equation}) are shown in Fig.~\ref{fig:comparision}-a as solid red ($n=0$) and dashed green ($n>0$) curves. 

\begin{figure*}[t]
  \vspace{-0.1in}
  \centering
  \includegraphics[width=\textwidth]{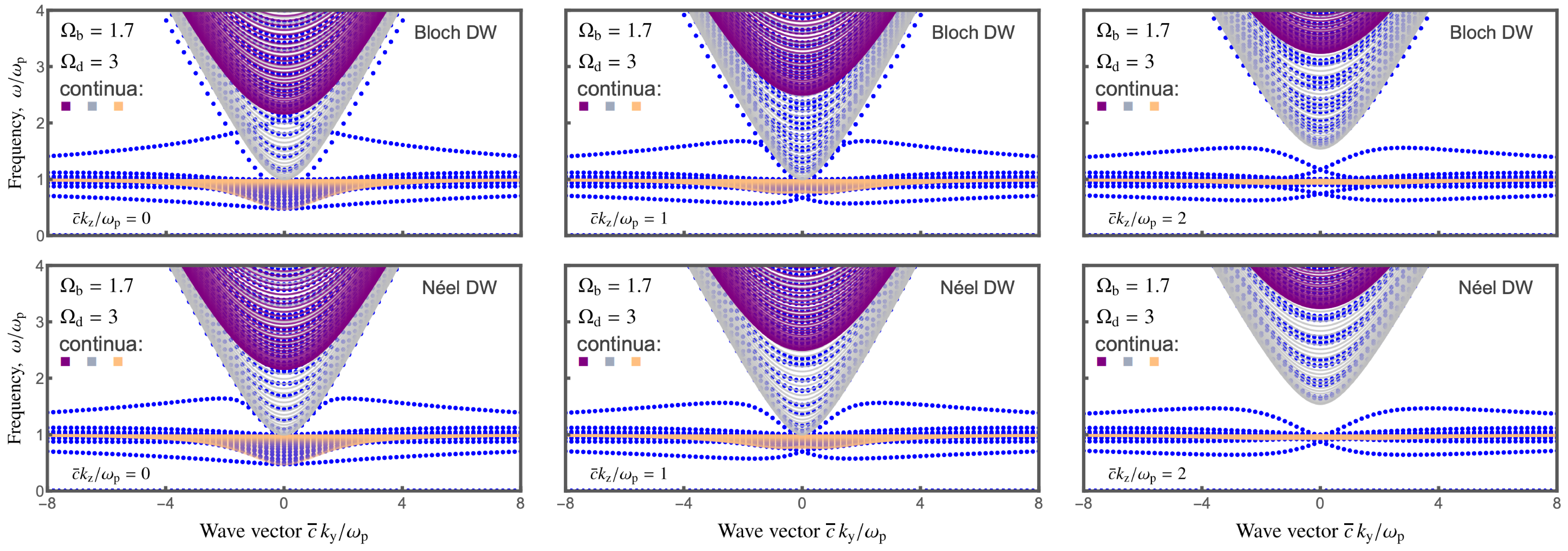}
  
\vspace{-0.15in}
  \caption{ The spectrum of the MP waves for Bloch (top row) and N\'eel (bottom row) DW
     profiles at $\bar{c}k_z/\omega_\mathrm{p}=0$ (left column), $1$ (middle column), and $2$ (right column). While the Kelvin- and Yanai-like modes exhibit distinct behaviour at small $\bar{c}k_z/\omega_\mathrm{p}$, their dispersions become qualitatively similar at intermediate and large wave vectors.}
  \vspace{-0.2in}
  \label{fig:bloch-neel}
\end{figure*}

The ground state of the PT problem gives rise to two chiral modes. One branch emerges from the bottom of the lowest-frequency continuum and re-enters it at large wavevectors. The second branch connects the two continua, reflecting its topological origin and closely resembling the Yanai plasmon mode of a two-dimensional electron gas confined by a magnetic-field kink. Because it is decoupled from oscillations of the $E_z$ component of the electric field, its dispersion freely traverses the corresponding continuum without being affected by it.

The excited states of the PT problem give rise to three distinct families of topologically trivial modes. Two families are clearly resolved, while the third emerges between the Yanai-like mode and the upper continuum only when the domain wall is sufficiently smooth, resembling Volkov-Pankratov-type electronic states~\cite{VP1,VP2,VP3} or plasmonic Poincare-like modes~\cite{EfimkinMP}. Although all three families originate from, and eventually merge back into, the same bulk continuum --- as expected for topologically trivial excitations --- only one of them exhibits chiral propagation. 

Analytical calculations are well reproduced numerically, with all numerically obtained interface states occurring in mirror-reflected pairs, consistent with the two symmetry-related domain walls in the profile $\vec{b}'(x)$ (See SM for details).

The calculated DW spectrum exhibits two topological branches (Kelvin-like and Yanai-like), consistent with the bulk--boundary correspondence arising from the interchange of Weyl nodes across the wall. Their protection, however, relies on the decoupling of the $E_z$ oscillations from the remaining field components, which occurs only in the Voigt configuration ($k_z = 0$) and when $b_x = b_y = 0$. These constraints are not satisfied for Bloch and N\'eel domain-wall profiles, and we assess the fate of the topological modes numerically.

 \noindent\textcolor{blue}{\emph{Numerical solution for the Bloch and N\'eel DW profiles}} The MP waves spectrum for Bloch (top row) and N\'eel (bottom) DW profiles is presented in Fig. \ref{fig:bloch-neel}. Three columns correspond to wave vector component values $\bar{c} k_z/\omega_\mathrm{p}=0, 1, 2$, with two of them between the MP Weyl nodes with magnitude  $\bar{c} k_\mathrm{W}/\omega_\mathrm{p}=\sqrt{2}$. 

For the Bloch DW profile, coupling between the Kelvin-like branch and the continuum of modes involving $E_z$ oscillations leads to a splitting of this mode and, therefore, its stabilisation. In contrast, the Yanai-like branch is only weakly affected by the continuum and undergoes noticeable modification only upon entering it. Both modes remain well defined for $\bar{c}k_z$ values between MP Weyl notes and continue to link the distinct continua, as expected for topological branches. Although true topological protection is lost due to the coexistence with the bulk modes in continua, the characteristic arc-like behaviour is nevertheless preserved.

For the N\'eel DW profile, the Kelvin-like branch remains embedded in the continuum of modes involving $E_z$ oscillations and is therefore not spectrally well resolved. The Yanai-like branch, by contrast, couples strongly to the continuum and is pushed away from it. Although the mode persists and remains strongly nonreciprocal, it no longer connects distinct continua. Its dispersion becomes qualitatively similar for wave vectors both between and outside the Weyl nodes, thereby losing the characteristic arc-like structure.

The distinct behaviour of the topological modes at small $\bar{c}k_z/\omega_\mathrm{p}$ can be traced to their different polarizations within the framework of perturbation theory (see SM for details). By contrast, this distinction gradually diminishes as $\bar{c}k_z/\omega_\mathrm{p}$ increases, and then the dispersions of the Kelvin and Yanai modes become very similar.

\noindent\textcolor{blue}{\emph{Discussion and conclusions.}} The family of magnetic Weyl semimetals is growing and now includes kagome ferro- and antiferromagnets (Co$_3$Sn$_2$S$_2$, Fe$_3$Sn$_2$), Heusler compounds (Co$_2$MnGa), and layered magnetic systems such as MnBi$_2$Te$_4$, etc~\cite{AHEMain1,AHEMain2,AHESecondary1,AHESecondary2,AHESecondary3}. For estimates, we focus on Co$_3$Sn$_2$S$_2$, which combines a giant anomalous Hall response with a relatively simple band structure hosting three inversion-related Weyl-node pairs and a well-characterised magnetic domain structure already observed using magneto-optical Kerr (MOKE) and magnetic force (MFM) microscopies~\cite{WeylDW1,WeylDW2}. The corresponding parameters include Fermi energy $\varepsilon_{\mathrm F}\approx 85\;\mathrm{meV}$, Weyl-node separation $b\approx 0.05\;\mathrm{\AA^{-1}}$, velocity $v\approx 3\times10^{7} \;\mathrm{cm/s}$, dielectric constant $\varepsilon_\infty\approx 10$, and DW width $d\approx 250~\mathrm{nm}$. They yield $\omega_{\mathrm p}\approx 82~\hbox{meV} $, $\omega_{\mathrm b}\approx 140~\hbox{meV}$, and $\omega_{\mathrm d}\approx 180~\hbox{meV} $, and the dimensionless parameters $\Omega_{\mathrm b}\approx 1.7$ and $\Omega_{\mathrm d}\approx 3$ used in the figures. The trapped modes are well developed over a broad energy range of $\sim 40\!-\!160~\mathrm{meV}$ ($10\!-\!40~\mathrm{THz}$), which are about an order of magnitude higher than in conventional magnetoplasma wave setups and are unachievable in laboratory-scale magnetic fields.

The most striking manifestation of the Weyl character of magnetoplasma waves is the emergence of domain-wall-confined modes with intrinsically nonreciprocal propagation. Once a magnetic domain wall on the surface of a magnetic Weyl semimetal is identified using real-space imaging techniques-such as MOKE or MFM~\cite{WeylDW1,WeylDW2} - the directional propagation of these modes can be probed using scattering-type scanning near-field optical microscopy (s-SNOM) or related near-field techniques developed for studying surface plasmon polaritons and edge magnetoplasmons in two-dimensional electronic systems~\cite{SNOMReview}. The predicted domain-wall modes closely resemble trapped electromagnetic states at other gyrotropic interfaces, including interfaces between a magnetised plasma and a perfect electric conductor or a transparent dielectric~\cite{GyrotropicFirst,GyrotropicInterfaces1,GyrotropicInterfacesReview}. In particular, they are closely related to the arc-like interface modes predicted at boundaries between magnetised plasmas with oppositely oriented gyration vectors~\cite{GyrotropicInterfacesMPTwist}.     

The interchange of electronic Weyl nodes across a magnetic DW ensures the presence of a single Fermi-arc state~\cite{FermiArc1,FermiArc2}, potentially accompanied by additional Volkov-Pankratov bound states if the wall is sufficiently smooth~\cite{FermiArcPlasmon3}. When the Fermi level intersects these arc bands, a qualitatively different trapped chiral plasmonic mode can emerge~\cite{FermiArcPlasmon1, FermiArcPlasmon2, FermiArcPlasmon3}. In the considered regime, however, all electronic domain-wall states lie far below the Fermi level and thus exert a negligible influence on the MP modes considered here.

In conclusion, we have shown that MP waves in magnetic Weyl semimetals exhibit a pair of Weyl nodes generated by the intrinsic anomalous Hall response. The nontrivial topology of this spectrum manifests itself through the emergence of unconventional domain-wall-confined modes, which could exhibit arc-like dispersion and intrinsically nonreciprocal propagation.

\textcolor{blue}{\emph{Acknowledgments}.} 
We acknowledge fruitful discussions with Haoran Ren and support from the Australian Research Council Centre of Excellence in Future Low-Energy Electronics Technologies (CE170100039) and from the Spanish Ministerio de Ciencia y Universidades-Agencia Estatal de Investigaci\'on through Grant No. PID2024-161142NB-I00. 

\bibliography{References}

\newpage

\appendix

\section{Axion electrodynamics perspective} 
This Appendix extends the effective-Hamiltonian reformulation developed in the main text to a more general scenario of a Weyl semimetal with intrinsic breaking of both inversion and time-reversal symmetries. Furthermore, we formulate the coupled dynamics of matter and electromagnetic fields within the framework of axionic electrodynamics. This approach provides a natural description of the topological anomalous Hall response through the axion field $\theta(\mathbf{r}, t)$ and an additional term in the electromagnetic Lagrangian density given by     
\begin{equation}
\mathcal{L}_{\theta}
= -\,\frac{e^{2}}{4\pi^{2}\hbar c}\,
\theta(\mathbf{r},t)\,\mathbf{E}\cdot\mathbf{H}
\label{axion action}
\end{equation}
For the model considered here --- a Weyl semimetal with $N$ identical pairs of Weyl nodes separated in momentum space by the vector $2\mathbf{b}$ and split in energy by $2b_0$ --- the axion field takes the form $\theta(\mathbf{r}, t) = 2N\big(\mathbf{b}\cdot\mathbf{r} - b_0 t\big)$. Its temporal and spatial derivatives determine two physically distinct topological contributions to electric current and are given by
\begin{equation}
\label{topological current}
\begin{split}
&\mathbf j_{\mathrm{C}}
= \frac{e^2}{4\pi^{2}\hbar c}\,(\partial_t \theta)\,\mathbf H = \frac{e^2}{2\pi^{2}\hbar c}\,b_0\,\mathbf H   \\
&\mathbf j_{\rm A}
= \frac{e^2}{4\pi^{2}\hbar}\,[\nabla \theta\times\mathbf E]=\frac{e^2}{2\pi^{2}\hbar}\,[\vec{b}\times\mathbf E].
\end{split}
\end{equation}
The first term describes an electric current flowing parallel to the magnetic field, associated with the chiral magnetic effect, and the second term corresponds to the topological anomalous Hall response. The least action principle, incorporating the axion term, results in the following coupled equations of matter and electromagnetic fields

\begin{subequations}
\begin{align}
\mathrm{curl}\,\mathbf{H}
&=
\frac{4\pi}{c}
\big(
\mathbf{j}_{\mathrm{D}}
+\mathbf{j}_{\mathrm{A}}
+\mathbf{j}_{\mathrm{C}}
\big)
+\frac{1}{c}
\frac{\partial \mathbf{D}}{\partial t},
\label{eq:axion_Maxwell1}
\\
\mathrm{curl}\,\mathbf{E}
&=
-\frac{1}{c}
\frac{\partial \mathbf{H}}{\partial t},
\label{eq:axion_Maxwell2}
\\
\intertext{Here, we have also incorporated the conventional matter response originating from the electric-field-induced shift of the electron distribution within the Fermi seas surrounding the Weyl nodes and described by the Drude model.}
\frac{\partial \mathbf{j}_{\mathrm{D}}}{\partial t}
&=
\mathcal{D}\,\mathbf{E}
+\frac{e}{m_\mathrm{c} c}\,
[(\mathbf{B}_0+\mathbf{H}) \times \mathbf{j}_{\mathrm{D}}],
\label{eq:axion_Maxwell3}
\end{align}
\end{subequations}
Here, $\mathcal{D}$ denotes the Drude weight, and $m_{\mathrm c}=\varepsilon_{\mathrm F}/v^{2}$ is the cyclotron mass of Weyl electrons. We also include a static external magnetic field $\vec{B}_0$. The contribution of the oscillating magnetic field $\mathbf{H}(\mathbf{r},t)$ to the Lorentz force vanishes upon linearization and is therefore unimportant here.
 
Similar to the procedure in the main text, Eqs.~\eqref{eq:axion_Maxwell1}--\eqref{eq:axion_Maxwell3}
can be rewritten as an eigenvalue problem for the rescaled nine-component state vector $\psi_\mathrm{pl} = \{\sqrt{\varepsilon_{\infty}}\vec{E}(\vec{k}),\, \vec{H}(\vec{k}),\, 4\pi\vec{j}_\mathrm{D}(\vec{k})/(\omega_{\mathrm{p}}\sqrt{\varepsilon_{\infty}})\}$. The corresponding effective Hamiltonian is given by
    \begin{equation}
\hat{\mathcal{H}}_\mathrm{pl}(\vec{k})=\begin{pmatrix}
-ic_b \hat{K}_\vec{b} & - \bar{c} \hat{K}_{\vec{k}}+i \omega_\mathrm{C} \hat{1} & -i\omega_\mathrm{p} \hat{1}\\
\bar{c} \hat{K}_{\vec{k}} & 0 & 0 \\
i \omega_\mathrm{p} \hat{1} & 0 &  i \omega_\mathrm{c} \hat{K}_{\mathrm{c}} \\
\end{pmatrix}
\label{full effective hamiltonian}
\end{equation}
Here, the antisymmetric matrix $\hat{K}_{\mathrm c}=\hat{K}_{\mathbf{B}_0/B_0}$
represents the cross-product operation associated with the Lorentz force induced by the external magnetic field $\mathbf{B}_0$. The frequencies $\omega_{\mathrm{C}}$ and $\omega_{\mathrm{C}}$ characterize the magnitudes of the chiral magnetoelectric and conventional Hall effects as 
\begin{equation}
\omega_{\mathrm{C}}=\frac{2e^2 b_0}{\pi \sqrt{\varepsilon_\infty}\hbar c}, \quad \quad \omega_c=\frac{e\mathbf{B}_0}{m_\mathrm{c} c}
 \end{equation}
Interestingly, in the presence of the chiral magnetic effect, the effective Hamiltonian is no longer Hermitian. Its eigenvalues, however, remain real, protected by the combined action of particle-hole and inversion symmetries. The topological classification of the general scenario and analysis of possible topological trapped modes are beyond the scope of the present Letter. 
\section{Chern number calculation}

\begin{figure}
\vspace{-0.05in}
\centering
\includegraphics[scale=0.5]{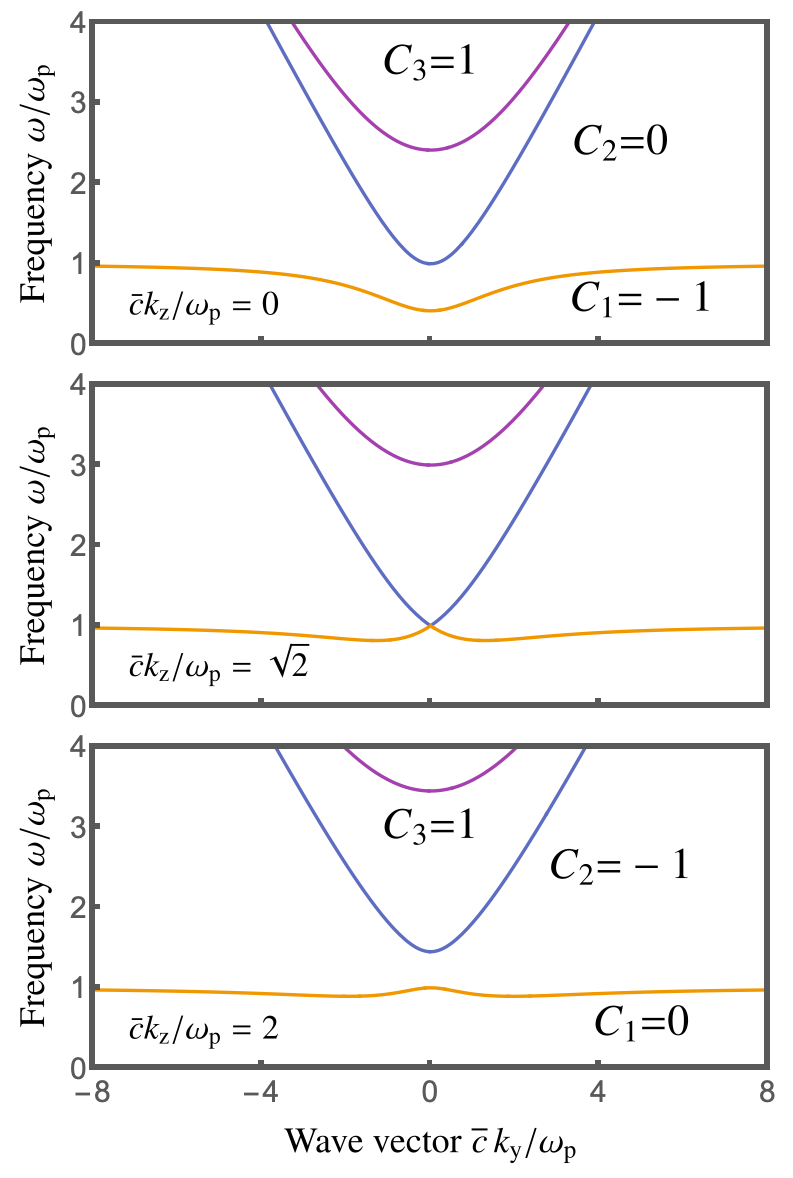}
\vspace{-0.06in}
\caption{The dispersion of magnetoplasma waves across reciprocal-space sections perpendicular to the gyrotropy axis, together with the corresponding topological Chern numbers. The exchange of Chern numbers at the crossing of two branches confirms their Weyl nature.} \vspace{-0.20in}\label{Bulk chern number}
\end{figure}

This Appendix provides details of the topological analysis of Weyl magnetoplasma waves. As in the case of conventional magnetoplasma waves supported by an electron gas in a magnetic field~\cite{PlasmaTop2}, a direct evaluation of the Chern numbers using Eq.~(\ref{Chern number}) yields noninteger values for the longitudinal (L) wave. This behavior has been recognized as an artifact of the noncompact reciprocal space inherent to coarse-grained hydrodynamic models that neglect the underlying crystalline lattice. Following Ref.~\cite{PlasmaTop2}, we introduce a regularization factor 
\begin{equation}
    r(k_x, k_y)
= \frac{1}{
1 + \left( \frac{\sqrt{k_x^2 + k_y^2}}{k_{\mathrm{max}}} \right)^{p}
},
\end{equation} 
as $\omega_\mathrm{p} \rightarrow \omega_\mathrm{p}\, r(k_x,k_y)$. Here $k_{\mathrm{max}}$ is a smooth reciprocal-space cutoff and $p>0$ is a regularization parameter. This procedure ensures that, for any $p>0$, the Berry curvature decays sufficiently rapidly at large wave vectors, rendering the Chern numbers of all branches integer-valued. The resulting set of Chern numbers for the three branches is $(1,\,0,\,-1)$ for reciprocal-space sections between the Weyl nodes, while for sections outside the nodes it becomes $(1,\,-1,\,0)$, as discussed in the main text and illustrated in Fig.~\ref{Bulk chern number}.

\section{The fate of domain-wall-trapped topological modes: a perturbative perspective}

\subsection{The perturbation theory framework}

In the main text, we discussed the distinct behaviour of topological Kelvin- and Yanai-like modes for Bloch and N\'eel domain-wall (DW) profiles, based on explicit numerical calculations. In this Appendix, we continue to assume the Voigt configuration ($k_z=0$), while gaining further insight into this behaviour by treating the non-$z$ components of the gyration vector --- crucial for the realistic Bloch and N\'eel DW profiles --- within the framework of quantum-mechanical perturbation theory. Really, the effective Hamiltonian can be presented as
\begin{equation}
\hat{\mathcal{H}}_\mathrm{pl}(\hat{k}_x,k_y,0)=\hat{\mathcal{H}}_\mathrm{pl}^0(\hat{k}_x,k_y,0)+\hat{\mathcal{H}}_\mathrm{pl}^{xy}(\hat{k}_x,k_y,0)
\end{equation}
The first $\hat{\mathcal{H}}_\mathrm{pl}^0(\hat{k}_x,k_y,0)$ corresponds to the simplified DW profile and, as discussed in the main text, admits the analytical solution. The second term, $\hat{\mathcal{H}}_\mathrm{pl}^{xy}(\hat{k}_x,k_y,0)$,  describes the anomalous Hall response induced due to non-z components of the gyration vector. Only its top-left  block $\hat{\mathcal{V}}$ is non-zero, and has the profile-dependent structure given by 
\begin{align}
\label{Bloch  perturbation matrix}
\text{Bloch:} \quad  \hat{\mathcal{V}}_\mathrm{B}=
\begin{pmatrix}
0 & 0 & -i\omega_{b}\\
0 & 0 & 0\\
i\omega_{b}& 0 & 0
\end{pmatrix} \frac{1}{\mathrm{cosh}(\frac{x}{d})}, \\ \text{Neel: } \quad 
\hat{\mathcal{V}}_\mathrm{N}=
\begin{pmatrix}
0 & 0 & 0\\
0 & 0 & i\omega_{b}\\
0 & -i\omega_{b} & 0
\end{pmatrix} \frac{1}{\mathrm{cosh}(\frac{x}{d})}.
\end{align}
We are interested only in the $k_y$ dependence of the topological Kelvin-like ($\alpha\equiv \mathrm{K}$) and Yanai-like ($\alpha\equiv \mathrm{Y}$) modes described by the states $|\alpha\rangle$. As evident from Figs.~(\ref{fig:comparision}) and (\ref{fig:bloch-neel}), the perturbation affects these modes only in the vicinity of, and within, the continuum of electric-field oscillations along the gyrotropy axis. The corresponding continuum modes can be labeled as $|E_z(k_x,k_y)\rangle$, with dispersion relation $\omega_{E_z}(k_y,k_x)=\sqrt{\omega_\mathrm{p}^2+\bar{c}^2(k_x^2+k_y^2)}$. Up to the second order in quantum-mechanical perturbation theory, the coupling between this continuum and the topological modes leads to a renormalization of the latter's frequencies as
\begin{equation}
\delta \omega_\alpha(k_y)=\langle\alpha| \hat{\mathcal{H}}_\mathrm{pl}^{xy}| \alpha\rangle + \sum_{k_x} \frac{|\langle E_z(k_x,k_y)| \hat{\mathcal{H}}_\mathrm{pl}^{xy}|\alpha \rangle|^2}{\omega_\alpha(k_y)-\omega_{E_z} (k_y,k_x)}
\end{equation}
The differing behaviour of $\delta\omega_\alpha(k_y)$ for the two modes and domain-wall profiles can be traced to the distinct structure of the corresponding matrix elements.
\subsection{Wave functions and matrix elements}
While all modes are described by nine-component vectors in the basis $\psi_\mathrm{pl}=\{\sqrt{\varepsilon_{\infty}}\vec{E}(\vec{k}),\,\vec{H}(\vec{k}),\,4\pi\vec{j}_\mathrm{D}(\vec{k})/(\omega_{\mathrm p}\sqrt{\varepsilon_{\infty}})\}$, not all components contribute equally. Since only the top-left $3\times 3$ block of $\hat{\mathcal{H}}_\mathrm{pl}^{xy}(\hat{k}_x,k_y,0)$ is nonzero, the relevant matrix elements are determined solely by the first three components of the vector, corresponding to oscillations of the electric field. Accordingly, the truncated vector describing $E_z$ oscillations within the continuum is given by \begin{equation}
\underline{|E_z(k_x,k_y)\rangle}\propto\begin{pmatrix}
0 \\
0 \\
1\\
\end{pmatrix}
e^{ik_x x}
\label{E_z oscillation}
\end{equation}
and describe a propagating plane wave. As for the Kelvin-like mode, its truncated vector is given by 
\begin{equation}
\underline{|\mathrm{K}(k_y)\rangle}\propto\begin{pmatrix}
1\\
0 \\
0\\
\end{pmatrix} \frac{1}{\mathrm{cosh}^\zeta\left(\frac{x}{d}\right)}
\end{equation}
where $\zeta=\sqrt{\omega_\mathrm{p}^2+\bar{c}^2k_y^2} \omega_\mathrm{b} d/\bar{c}^2k_y$. This trapped mode involves oscillations of the $E_x$ component of the electric field only, and its spatial profile is even. The Yanai-like mode has different polarization governed by the following truncated vector 
\begin{equation}
    \underline{|\mathrm{Y}(k_y)\rangle}\propto\begin{pmatrix}
A \tanh(\frac{x}{d})\\
1\\
0\end{pmatrix}
 \frac{1}{\mathrm{cosh}^\lambda(\frac{x}{d})}
\end{equation}
Here, $\lambda$ is the P\"oschl--Teller (PT) parameter defined in the main text, and $A$ is a $k_y$-dependent constant. Its explicit, cumbersome form is not important here. What matters is that this mode involves oscillations of both field components, $E_x$ and $E_y$, whose spatial distributions are odd and even, respectively.
First, we note that the first-order frequency corrections
\begin{equation}
\langle\alpha|\hat{\mathcal{H}}_\mathrm{pl}^{xy}|\alpha\rangle
=\underline{\langle\alpha|}\,\hat{\mathcal{V}}_\alpha\,\underline{|\alpha\rangle}
=0
\end{equation}
vanish for both topological modes and for both domain-wall profiles. By contrast, the second-order corrections are nonzero. The corresponding matrix elements for the Kelvin-like mode are given by 

\begin{align}
&\underline{\langle E_z(k_x,k_y)|} \hat{\mathcal{V}}_\mathrm{B}\underline{|\mathrm{K} \rangle} \propto 
\int_{-\infty}^{+\infty} dx\; \frac{1}{\mathrm{cosh}^{\zeta+1}\left(\frac{x}{d}\right)} e^{i k x} \\
&\underline{\langle E_z(k_x,k_y)|} \hat{\mathcal{V}}_\mathrm{N}\underline{|\mathrm{K}\rangle} \propto  0
\end{align}
For the Bloch DW profile, the coupling between the topological Kelvin-like mode and propagating $E_z$ oscillations is finite and maximal in the long-wavelength limit. This coupling induces a redshift of the mode and leads to its spectral stabilization. In contrast, for the N\'eel DW profile, the corresponding overlap vanishes, so the Kelvin-like branch remains embedded in the bulk continuum associated with $E_z$ oscillations and is not spectrally resolved.

The corresponding matrix elements for the Yanai-like mode are given by 
\begin{align}
&\underline{\langle E_z(k_x,k_y)|} \hat{\mathcal{V}}_\mathrm{B}\underline{|\mathrm{Y} \rangle} \propto \int_{-\infty}^{+\infty} dx\;
\frac{ \tanh(\frac{x}{d}) }{\cosh^{\lambda+1}(\frac{x}{d})} e^{i k_x x} \\
&\underline{\langle E_z(k_x,k_y)|} \hat{\mathcal{V}}_\mathrm{N}\underline{|\mathrm{Y} \rangle} \propto \int_{-\infty}^{+\infty} dx\;
\frac{ 1}{\cosh^{\lambda+1}(\frac{x}{d})}e^{i k_x x}
\end{align}
In the long-wavelength limit $k_x \to 0$, the matrix element for the Bloch DW profile vanishes due to the odd parity of the integrand. As a result, the coupling between the Yanai-like mode and the bulk continuum associated with $E_z$ oscillations remains weak and becomes appreciable only once this topological mode penetrates the continuum. In contrast, for the N\'eel DW profile the matrix element is finite, and the mode frequency is effectively repelled from the continuum, thereby preventing its penetration.

The above analysis based on quantum mechanical perturbation theory is in excellent agreement with the results obtained from the explicit numerical solutions presented in the main text. 

\vspace{0.20in}
\section{The spectrum of MP waves for sharp and smooth domain wall profiles}
\begin{figure}[b]
  \centering
 \includegraphics[width=0.95\linewidth]{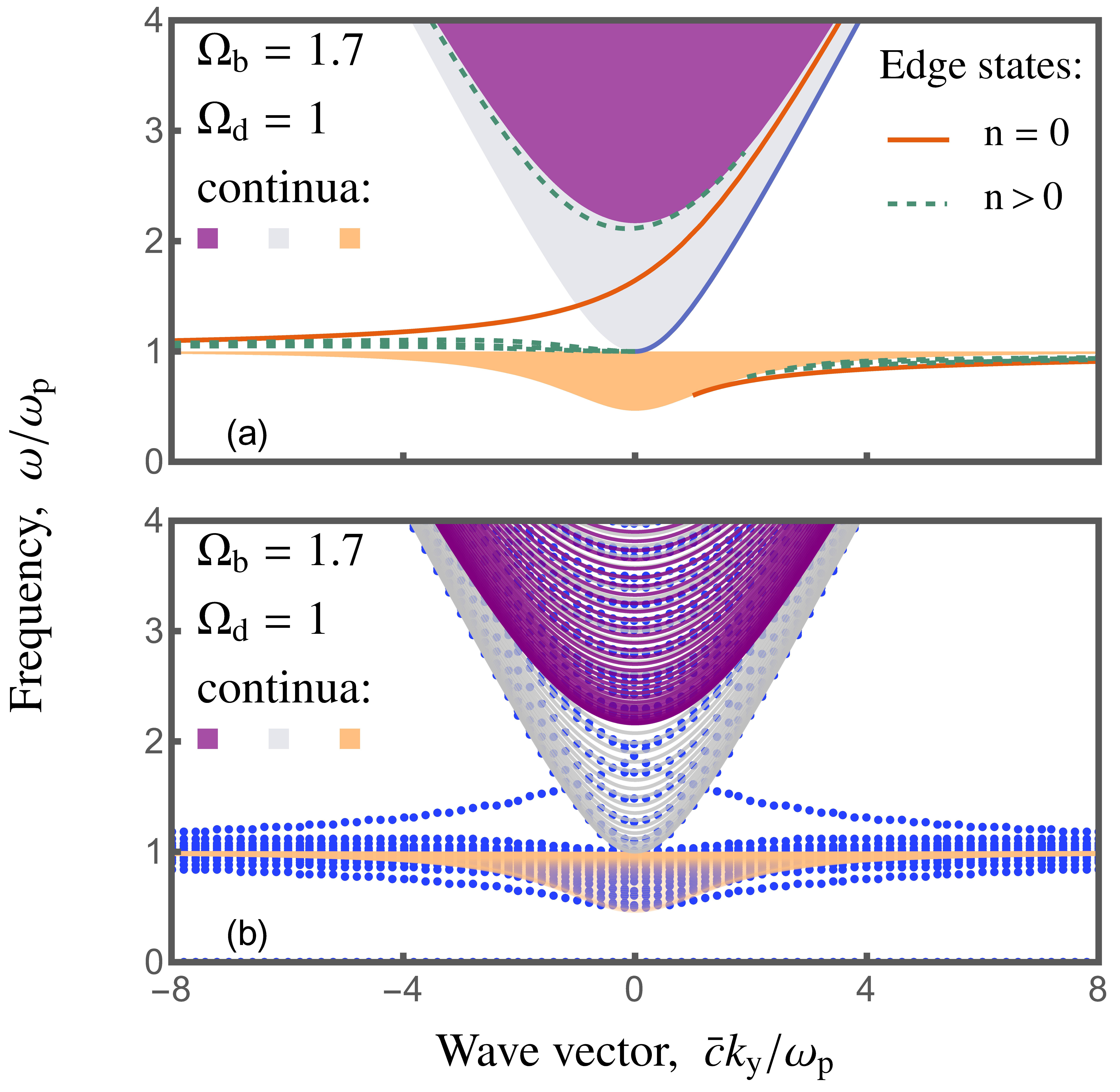}
  \caption{Spectrum of bulk (shaded regions) and DW-confined (lines) MP waves for the simplified profile in Eq.~(\ref{simplifiedDW}) at $k_z = 0$, calculated analytically (a) and numerically (b). }
  \label{smoothDW}
\end{figure}

This Appendix presents analytical and numerical results for the smoother DW profile, corresponding to $\Omega_\mathrm{d}=1$, as shown in Fig.~\ref{smoothDW}. For the set of parameters corresponding to $\mathrm{Co_3Sn_2S_2}$, it corresponds to a width $d \approx 770\,\mathrm{nm}$. The spectrum exhibits an additional topologically trivial mode lying within the continuum of $E_z$-oscillations and resembling Volkov--Pankratov--type electronic states~\cite{VP1,VP2,VP3} or plasmonic Poincar\'e-like modes~\cite{EfimkinMP}. The number of modes in this set increases as the domain-wall width increases. 

Analytical calculations are well reproduced numerically, with all numerically obtained interface states occurring in mirror-reflected pairs, consistent with the two symmetry-related domain walls in the profile $\vec{b}'(x)$.

\end{document}